\documentclass[11pt,a4paper,oneside]{article}
\usepackage{lineno,hyperref}

\usepackage{graphicx}              
\usepackage{epsfig}
\usepackage{amsmath}               
\usepackage{amssymb}
\usepackage{epstopdf}
\usepackage{verbatim}
\usepackage{mathalpha}
\usepackage{mathtools}
\usepackage[scale=0.80]{geometry}
\usepackage{graphicx, times}
\usepackage{amsfonts}              
\usepackage{amsthm}                
\usepackage{multicol}
\usepackage{algorithm}
\usepackage{algorithmic}
\usepackage{varwidth}
\usepackage{parskip}
\usepackage{hyperref}
\usepackage{rotating}
\usepackage[numbers,sort&compress]{natbib}
\usepackage{multirow}
\usepackage{pdflscape}
\usepackage[numbers]{natbib}
\usepackage{caption}
\usepackage{xcolor}
\usepackage{color}
\usepackage{comment}
\usepackage{subcaption}
\newcommand*{\rom}[1]{\expandafter\@\romannumeral #1}

\newcommand{\bea}{\begin{eqnarray}}
	\newcommand{\eea}{\end{eqnarray}}
\newcommand{\bee}{\begin{eqnarray*}}
	\newcommand{\eee}{\end{eqnarray*}}

\begin{document}
\author{Khomesh R. Patle$^{1}$\footnote{khomeshpatle5@gmail.com} and G. P. Singh$^{1}$\footnote{gpsingh@mth.vnit.ac.in}
\vspace{.2cm}\\
${}^{1}$ Department of Mathematics,\\ Visvesvaraya National Institute of Technology, Nagpur, 440010, Maharashtra, India.
\vspace{.2cm}
\date{}}
\title{A new parametric study of $f(T)$ teleparallel gravity with generalized Chaplygin gas: Confronting observational data}
\maketitle
\begin{abstract} 
$\hspace*{0.1cm}$ In this paper, we investigate the cosmological dynamics of a modified gravity framework based on the function $f(T)=\beta(-T)^{1/2}+\gamma(-T)$, where $T$ denotes the torsion scalar. The matter sector is modeled using the Generalized Chaplygin Gas (GCG) with the equation of state $p=-\frac{\mathcal{A}}{\rho^{\eta}}$, allowing the model to describe the evolution from an early-time matter-like phase to a late-time accelerated universe. By deriving an analytical expression for the Hubble parameter $H(z)$, we perform parameter estimation using Bayesian statistical techniques based on the $\chi^{2}$-minimization method with the Cosmic Chronometer (CC) and joint (CC+Pantheon) observational datasets. The deceleration parameter exhibits a transition from deceleration to acceleration, while the present-day value indicates the current accelerated expansion. The model yields a present-day EoS parameter consistent with dark-energy-like behavior, while the energy conditions indicate that the NEC and DEC are satisfied and the SEC is violated at late times. The $\omega-\omega'$ plane shows a freezing quintessence behavior, approaching the $\Lambda$CDM-like point $(\omega,\omega')=(-1,0)$, while the cosmographic analysis offers additional insight into the evolution of the cosmic expansion. The estimated age of the universe is also consistent with current observational bounds. These results demonstrate that the GCG scenario within $f(T)$ teleparallel gravity provides a viable and observationally consistent framework for explaining the late-time accelerated expansion of the universe.
\end{abstract}
\section{Introduction}\label{sec:1} 
$\hspace*{0.5cm}$ At the end of the twentieth century, a series of observational discoveries provided compelling evidence that the universe is undergoing accelerated expansion~\cite{1998AJ....116.1009R,1999ApJ...517..565P,2020A&A...641A...6P}. This unexpected phenomenon led to the proposal of a new form of cosmic energy, commonly referred to as dark-energy (DE), which is invoked to account for the observed acceleration. Despite its dominant contribution to the present-day energy budget of the universe, the fundamental nature and physical origin of DE remain poorly understood, making it one of the most important open questions in modern cosmology. Current observations suggest that dark-matter and dark-energy together account for approximately $95$–$96\%$ of the total cosmic energy density~\cite{weinberg1989cosmological}. Among the various possibilities proposed to describe DE, the cosmological constant ($\Lambda$) provides the simplest and most successful phenomenological framework.
\vspace{0.2cm}\\
$\hspace*{0.5cm}$ The $\Lambda$CDM model has demonstrated remarkable consistency with a broad range of cosmological observations and is therefore regarded as the standard model of modern cosmology. Nevertheless, the model is accompanied by fundamental theoretical difficulties, most notably the fine-tuning and cosmic coincidence problems~\cite{di2021realm,carroll2001cosmological,padmanabhan2003cosmological,copeland2006dynamics}. These unresolved issues have motivated considerable efforts to develop alternative descriptions of the late-time expansion of the universe. In particular, two broad approaches have been extensively investigated: introducing new forms of matter or energy with non-standard properties and modifying the gravitational framework itself. The latter possibility has attracted significant interest, as modified theories of gravity can account for the observed cosmic acceleration through geometric modifications of the gravitational sector, thereby reducing or eliminating the need to introduce a separate dark-energy component. Motivated by this perspective, numerous modified gravity theories have been proposed and investigated in the context of cosmology. These include a wide range of extensions and generalizations of General Relativity with applications to the background evolution, structure formation, and late-time accelerated expansion of the universe~\cite{buchdahl1970non,harko2011f,nojiri2011unified,jimenez2018coincident,nojiri2017modified,bamba2010finite,elizalde2010lambdacdm,harko2010f,capozziello2019extended,capozziello2023role,kotambkar2017anisotropic,hulke2020variable,singh2025observational,escamilla2024exploring,shukla2025multi,patle2026accelerated,singh2024conservative,Koussour,goswami2024flrw}.
\vspace{0.2cm}\\
$\hspace*{0.5cm}$ Among the different modified gravity theories developed to address the origin of cosmic acceleration, $f(T)$ teleparallel gravity~\cite{Bengochea,cai2016f} has emerged as an important and extensively investigated framework. The theory is constructed as a generalization of the teleparallel equivalent of General Relativity (TEGR), in which the gravitational Lagrangian is extended from the torsion scalar $T$ to a general function $f(T)$. In this formulation, the gravitational dynamics are expressed in terms of torsion rather than spacetime curvature. Although the construction of $f(T)$ gravity is formally analogous to that of $f(R)$ gravity, where the Ricci scalar $R$ in the Einstein--Hilbert action is replaced by a general function $f(R)$, their underlying geometrical descriptions are distinct. In General Relativity (GR), gravity is characterized by the curvature of spacetime through the torsion-free Levi--Civita connection, whereas teleparallel gravity employs a curvature-free Weitzenböck connection, with torsion serving as the fundamental geometrical quantity. For the particular case of TEGR, the torsion-based formulation reproduces the gravitational dynamics of GR. However, replacing $T$ by a general function $f(T)$ leads to a modified gravitational theory with dynamics that differ from those of GR and can produce a rich phenomenology at the cosmological level. In particular, $f(T)$ gravity provides a geometrically motivated framework for investigating the late-time accelerated expansion of the universe without necessarily introducing a separate dark-energy component. Its comparatively simple field equations, together with its broad cosmological applications, have consequently made $f(T)$ gravity an active area of research.
\vspace{0.2cm}\\
$\hspace*{0.5cm}$ The cosmological behavior of $f(T)$ gravity has been investigated extensively from different theoretical and observational perspectives. Previous studies have examined the dynamical evolution of the universe~\cite{paliathanasis2016cosmological}, thermodynamic properties~\cite{salako2013lambdacdm}, cosmographic reconstruction~\cite{capozziello2011cosmography}, and energy conditions~\cite{liu2012energy}. The theory has also been applied to early-universe scenarios such as matter-bounce cosmology~\cite{cai2011matter}. From the observational perspective, Zhadyranova et al.~\cite{zhadyranova2024exploring} constrained a linear $f(T)$ model using observational data to study the late-time accelerated expansion. Different functional forms of $f(T)$, Noether symmetries and model-independent methods for solving the cosmological field equations have also been studied~\cite{bamba2011equation,paliathanasis2014new,capozziello2017model}. A detailed review of the theoretical and cosmological aspects of $f(T)$ gravity can be found in Ref.~\cite{cai2016f}. Several recent works have further explored its cosmological applications~\cite{duchaniya2024attractor,koussour2024exploring,maurya2023anisotropic,bamba2016bounce,kavya2024can,bhar2024anisotropic,nunes2016new,duchaniya2022dynamical,chakraborty2023classical,dixit2021probe,patle2026,ren2022gaussian,chen2024prospects}. These investigations demonstrate the broad applicability of $f(T)$ gravity in describing the dynamics and evolution of the universe.
\vspace{0.2cm}\\
$\hspace*{0.5cm}$ In the search for a unified description of dark-matter and dark-energy, the Chaplygin gas (CG) model has emerged as an interesting candidate. The original Chaplygin gas, motivated by higher-dimensional theories such as string theory and brane-world scenarios, is characterized by the equation of state $p=-\frac{\mathcal{A}}{\rho}$, where $\mathcal{A}>0$ is a constant~\cite{Kamenshchik2001}. The model exhibits a remarkable dual behavior during cosmic evolution: it behaves like pressureless matter in the early universe and gradually develops a negative pressure at late times, resulting in an accelerated expansion. This characteristic makes the Chaplygin gas a natural candidate for describing the dark-sector through a single cosmic fluid. Recognizing the limitations of the original Chaplygin gas model in providing a sufficiently flexible description of cosmological observations, the Generalized Chaplygin Gas (GCG) was introduced by extending its equation of state to $p=-\frac{\mathcal{A}}{\rho^{\eta}}$~\cite{Bento2002}. Here, $\eta$ is an additional parameter that determines the evolution of the fluid and regulates the transition between its matter-like and dark-energy-like behaviors. The GCG therefore provides a more general framework in which the cosmic fluid evolves from a matter-dominated regime at early times toward a dark-energy-dominated phase at late times. In this way, the model can effectively reproduce the principal features of cold dark-matter and the cosmological constant within a unified description of the dark-sector. The incorporation of the GCG into $f(T)$ teleparallel gravity provides a further generalization of this unified dark-sector scenario. In $f(T)$ gravity, the gravitational dynamics are modified by replacing the torsion scalar $T$ in the teleparallel action with a general function $f(T)$. This modification changes the cosmological field equations and consequently affects the evolution of the cosmic fluid and the expansion history of the universe. The combination of the GCG with $f(T)$ gravity can therefore lead to modified behaviors of important cosmological quantities, such as the Hubble parameter, deceleration parameter and effective equation of state parameter. Furthermore, the additional freedom associated with the choice of the function $f(T)$ may provide a broader framework for describing the transition from decelerated to accelerated expansion and for confronting the model with observational data. Moreover, the persistent discrepancy between the early- and late-universe determinations of the Hubble constant, commonly known as the Hubble tension~\cite{di2021realm}, has further motivated the investigation of alternative cosmological models and modified gravity theories. In this context, the GCG model within $f(T)$ gravity provides a useful framework for constraining the Hubble parameter using observational data. Thus, studying the GCG in the context of $f(T)$ gravity offers an interesting possibility for investigating the unified dark-sector and the late-time dynamics of the universe.
\vspace{0.2cm}\\
$\hspace*{0.5cm}$ The main objective of this work is to explore the cosmological implications of the Generalized Chaplygin Gas within the framework of $f(T)$ teleparallel gravity. By combining the unified dark-sector description of the GCG with the modified gravitational dynamics of $f(T)$ gravity, we study the late-time evolution of the universe within this framework. The model parameters are constrained using observational data from the cosmic chronometer (CC) dataset and the joint (CC+Pantheon) dataset. The resulting cosmological behavior is further examined through relevant physical and cosmographic parameters to assess the consistency of the model with the observed expansion history. Overall, this study aims to assess the viability of the GCG scenario in $f(T)$ gravity and its ability to describe the observed late-time accelerated expansion of the universe.
\vspace{0.2cm}\\
\hspace*{0.5cm} The paper is organized as follows: In section~(\ref{sec:2}), we present the field equations of $f(T)$ teleparallel gravity together with the selected $f(T)$ model and discusses the resulting cosmological dynamics. Section~(\ref{sec:3}) focuses on the Generalized Chaplygin Gas (GCG) model, where we derive the corresponding $H(z)$ expression for our model. In section~(\ref{sec:4}), we describe the observational datasets employed to constrain the model parameters. Section~(\ref{sec:5}) is devoted to the evolution of the relevant cosmological parameters using the constrained model parameters. The standard energy conditions are analyzed in section~(\ref{sec:6}), followed by the dynamical behavior in the $(\omega-\omega')$ plane in section~(\ref{sec:7}). Section~(\ref{sec:8}) presents the cosmographic analysis based on the jerk and snap parameters, while the cosmic age is determined in section~(\ref{sec:9}). Finally, in section~(\ref{sec:10}), we summarize the principal findings and conclude the study.
\section{Field equations and cosmological dynamics in $f(T)$ teleparallel gravity}\label{sec:2}
$\hspace*{0.5cm}$ This section introduces the theoretical framework of $f(T)$ gravity and establishes the basic equations required for the subsequent cosmological analysis. In $f(T)$ gravity, the gravitational interaction is formulated through the torsion scalar $T$, while the modification of the gravitational dynamics is encoded in a general function $f(T)$. The spacetime geometry is described by the following line element:
\begin{equation}{\label{1}}
ds^{2}= g_{\mu\nu}dx^{\mu}dx^{\nu}= \eta_{lm}\theta^{l}\theta^{m},
\end{equation}
along with the associated components
\begin{equation}{\label{2}}
dx^{\mu} = e^{\mu}_{l}\theta^{l},~~~~~  \theta^{l}= e^{l}_{\mu} dx^{\mu},
\end{equation}
where $\eta_{lm}$ = $\mathrm{diag}(-1,1,1,1)$ corresponds to the Minkowskian metric, and $\left\{e^{l}_{\mu}\right\}$ denotes the tetrad field components. These tetrad fields are required to satisfy the following relations:
\begin{equation}{\label{3}}
e^{~~\mu}_{l} e^{l}_{~~\nu}= \delta^{\mu}_{\nu},~~~~~  e^{~~l}_{\mu} e^{\mu}_{~~m}= \delta^{l}_{m}.
\end{equation}
The geometric structure of $f(T)$ gravity is formulated in terms of the Weitzenböck connection~\cite{aldrovandi2012teleparallel}, whose definition is given by
\begin{equation}{\label{4}}
	\Gamma^{\alpha}_{\mu \nu}= e_{l}^{~\alpha} \partial_{\mu} e^{l}_{~\nu}= -e^{l}_{~\mu}\partial_{\nu} e^{~\alpha}_{l}.
\end{equation}
Corresponding to the above connection, the torsion tensor takes the following form~\cite{linder2010einstein}:
\begin{equation}{\label{5}}
T^{\alpha}_{~\mu \nu} = -\left(\Gamma^{\alpha}_{\nu \mu}-\Gamma^{\alpha}_{\mu \nu}\right)= - e^{~\alpha}_{l} \left(\partial_{\mu} e^{l}_{~\nu} - \partial_{\nu} e^{l}_{~\mu}\right).
\end{equation}
The torsion tensor provides the basis for defining the associated contorsion tensor, given by
\begin{equation}{\label{6}}
K^{\mu \nu}_{~\alpha} = -\frac{1}{2} \left(T^{\mu \nu}_{~\alpha} - T^{\nu \mu}_{\alpha} - T^{~\mu \nu}_{\alpha}\right),
\end{equation}
which, together with the torsion tensor, leads to the definition of the superpotential tensor
\begin{equation}{\label{7}}
S^{~\mu \nu}_{\alpha} = \frac{1}{2} \left(K^{\mu \nu}_{\alpha} + \delta^{\mu}_{\alpha} T^{\lambda \nu}_{~\lambda} - \delta^{\nu}_{\alpha}T^{\lambda \mu}_{~\lambda}\right).
\end{equation}
Using the torsion tensor and the superpotential $S^{~\mu \nu}_{\alpha}$, the torsion scalar $T$ is defined as follows~\cite{cai2016f,maluf2013teleparallel}:
\begin{equation}{\label{8}}
T= S^{~\mu \nu}_{\alpha} T^{\alpha}_{~\mu \nu} = \frac{1}{2}T^{\alpha \mu \nu } T_{\alpha \mu \nu} + \frac{1}{2}T^{\alpha \mu \nu } T_{\nu \mu \alpha} - T^{~\alpha}_{\alpha \mu } T^{\nu \mu}_{~\nu}.
\end{equation} 
The gravitational dynamics of the theory are governed by the following action~\cite{Bengochea,koussour2024exploring}:
\begin{equation}{\label{9}}  
	S=  \frac{1}{2\kappa^{2}}\int d^{4}xe \left[T+f(T)\right] + \int d^{4}xe L_{m},
\end{equation}
here, $e$ represents the determinant of the tetrad field, which is related to the metric determinant through $e=\det(e^{l}_{~\mu})=\sqrt{-g}$. The corresponding field equations are obtained by varying the action~(\ref{9}) with respect to the tetrad fields, yielding
\begin{equation}{\label{10}}
S^{~\nu \rho}_{\mu} \partial_{\rho} T f_{TT} + [e^{-1} e^{l}_{\mu} \partial_{\rho} (ee^{~\mu}_{l} S^{~\nu \lambda}_{\alpha}) + T^{\alpha}_{~\lambda \mu} S^{~\nu \lambda}_{\alpha}] f_{T} + \frac{1}{4}\delta^{\nu}_{\mu} f = \frac{\kappa^{2}}{2} \mathit{T}_{\mu}^{\nu},
\end{equation}
where $f_{T}= \frac{\partial f}{\partial T}$ and $f_{TT}= \frac{\partial^{2} f}{\partial T^{2}}$ denote the first and second derivatives of $f(T)$ with respect to the torsion scalar, respectively, while $\mathit{T}_{\mu}^{\nu}$ represents the energy-momentum tensor, which is given by
\begin{equation}{\label{11}}
	\mathit{T}_{\mu}^{\nu} = (\rho + \mathit{p}) u_{\mu} u^{\nu} + p \delta^{\nu}_{\mu},
\end{equation}
where $\rho$ and $p$ represent the energy density and pressure of the cosmic fluid, respectively. The four-velocity $u^{\mu}$ is normalized according to the condition $u^{\mu}u_{\mu}=-1$.
\vspace{0.1cm}\\
$\hspace*{0.5cm}$ For the cosmological analysis, we adopt a spatially flat Friedmann–Lemaître–Robertson–Walker (FLRW) spacetime, which provides a suitable background for studying the implications of $f(T)$ gravity. Under this assumption, the gravitational field equations can be reduced to the corresponding modified Friedmann equations. The line element for a spatially flat FLRW universe is given by~\cite{linder2010einstein}:
\begin{equation}{\label{12}}
	ds^{2}=-dt^{2}+a^{2}(t) \delta_{lm} dx^{l} dx^{m},
\end{equation}
here $a(t)$ represents the scale factor. For the line element given in Eq.~(\ref{12}), the torsion scalar is obtained as $T=-6H^{2}$.
\vspace{.1cm}\\
For the metric specified in Eq.~(\ref{12}), the modified Friedmann equations are obtained as~\cite{cai2016f}:
\begin{equation}{\label{13}}
2 \kappa^{2}\rho=6H^{2}+ 12H^{2}f_{T}+f,
\end{equation}
\begin{equation}{\label{14}}
-2 \kappa^{2}p=2\left(2\dot{H}+3H^{2}\right)+f+4 \left(\dot{H}+3H^{2}\right) f_{T}-48H^{2}\dot{H}f_{TT}.
\end{equation}
In the above equations, an overdot denotes differentiation with respect to cosmic time ($t$) and $H$ represents the Hubble parameter. The quantities $\rho$ and $p$ denote the total energy density and pressure of the cosmic fluid, respectively. We adopt the natural units $\kappa^{2}=1$~\cite{patle2026} throughout the subsequent analysis. The function $f(T)$ and its derivatives describe the modifications to the gravitational dynamics arising from the torsion-based formulation of gravity. The resulting equations provide a modified description of the cosmic expansion and allow the torsion-induced modifications to influence the late-time accelerated expansion of the universe.
\vspace{0.1cm}\\
$\hspace*{0.5cm}$ To facilitate a comprehensive cosmological analysis, we adopt a specific functional form of the gravitational Lagrangian in the $f(T)$ framework, given by
\begin{equation}{\label{15}}
f(T)=\beta(-T)^{\frac{1}{2}}+ \gamma(-T),
\end{equation}
where $\beta$ and $\gamma$ are constant model parameters associated with the respective torsional contributions. The adopted form incorporates both square-root and linear terms in the torsion scalar and provides a simple extension of the teleparallel gravitational framework. This choice enables us to investigate how the modified torsional dynamics affect the cosmic expansion history and the late-time accelerated phase of the universe.
\vspace{0.1cm}\\
$\hspace*{0.5cm}$ Several functional forms of $f(T)$ have been proposed to explore modifications of the teleparallel description of gravity and their implications for cosmic evolution. For instance, power-law models of the form $f(T)\propto(-T)^n$ have been investigated to describe deviations from the standard teleparallel framework~\cite{koussour2024exploring,Chakrabortty2023}. Exponential and logarithmic forms of $f(T)$ have also been considered as alternative functional choices for studying the late-time expansion and observational viability of modified teleparallel cosmologies~\cite{bamba2011equation}. These studies indicate that the functional dependence of the gravitational Lagrangian on the torsion scalar provides an useful way to explore departures from the standard cosmological dynamics. Motivated by these developments, we consider a mixed functional form containing square-root and linear dependences on $T$, providing a simple and tractable extension of the teleparallel gravitational Lagrangian for examining its cosmological implications within this framework.
\vspace{0.1cm}\\
$\hspace*{0.5cm}$ Upon implementing the proposed $f(T)$ form in the cosmological field equations, a noteworthy feature of the resulting background dynamics emerges. For the spatially flat FLRW background considered in this work, the contribution associated with the parameter $\beta$ cancels from the resulting background equations. Similar situations, in which an additional term in a generalized gravitational Lagrangian does not provide an independent contribution to the field equations, have been discussed in other modified gravity frameworks. For instance, in $f(T,B)$ teleparallel gravity, a linear dependence on the boundary term $B$ does not introduce an independent contribution to the field equations, owing to the boundary nature of $B$~\cite{Bahamonde2017}. A related feature also appears in metric-affine $F(R,\mathcal{D})$ gravity, where the term linear in $\mathcal{D}$ does not contribute to the field equations because its contribution to the action is a total divergence~\cite{Kenzhalin2024}. These examples illustrate that the presence of an additional term in a generalized gravitational function does not necessarily imply an independent contribution to the background dynamics. In the present model, consequently, the background cosmological evolution is governed by the remaining $\gamma$-dependent contribution, while $\beta$ does not provide an independent contribution to the background dynamics. With this gravitational framework established, we next introduce the generalized Chaplygin gas (GCG) and investigate its cosmological dynamics within the $f(T)$ framework.
\section{Generalized Chaplygin Gas (GCG) model in modified gravity}\label{sec:3}
\hspace*{0.5cm} We study the cosmic evolution in the $f(T)$ gravitational framework by considering the universe to be filled with a Chaplygin-type cosmic fluid. The Chaplygin gas (CG), initially introduced in the context of fluid dynamics, was later proposed as a unified description of the dark-matter and dark-energy sectors~\cite{Kamenshchik2001}. Its equation of state is given by
\begin{equation}{\label{16}}
	p=-\frac{\mathcal{A}}{\rho},
\end{equation}
where $\mathcal{A}>0$ is a constant. At early times, the fluid exhibits matter-like behavior with $p\approx0$, whereas at late times it approaches a cosmological-constant-like state characterized by $p\approx-\rho$.
\vspace{0.2cm}\\
\hspace*{0.5cm} To provide a more suitable description in light of observational studies, the Chaplygin gas model was extended to the generalized Chaplygin gas (GCG)~\cite{Bento2002}. The corresponding equation of state takes the form
\begin{equation}{\label{17}}
	p=-\frac{\mathcal{A}}{\rho^{\eta}},
\end{equation}
where $\mathcal{A}$ and $\eta$ are model parameters. The standard Chaplygin gas is recovered for $\eta=1$, while $\eta=0$ gives a constant negative pressure, corresponding to a cosmological-constant-like behavior. For a homogeneous and isotropic universe, the GCG fluid obeys the standard conservation equation:
\begin{equation}{\label{18}}
	\dot{\rho}+3H(\rho+p)=0.
\end{equation}
By inserting the GCG equation of state given in Eq.~(\ref{17}) into the conservation equation (\ref{18}), we obtain the corresponding evolution of the energy density as
\begin{equation}{\label{19}}
	\rho(z) = \left[\mathcal{A}+ C(1+z)^{3(1+\eta)}\right]^{\frac{1}{(1+\eta)}},
\end{equation}
where $C$ is an integration constant, and we have used the standard relation between the scale factor and redshift, $a=(1+z)^{-1}$. It is useful to express Eq.~(\ref{19}) in terms of the normalized parameter $\mathcal{A}_{c}$, defined by $\mathcal{A}_{c}=\frac{\mathcal{A}}{\rho_{0}^{1+\eta}}$, where $\rho_{0}$ denotes the present-day energy density. In terms of this parameter, Eq.~(\ref{19}) can be written as
\begin{equation}{\label{20}}
	\rho(z) = \rho_{0} \left[\mathcal{A}_{c}+ (1-\mathcal{A}_{c})(1+z)^{3(1+\eta)}\right]^{\frac{1}{(1+\eta)}}.
\end{equation}
This form is well suited for cosmological and observational analyses, as it offers a convenient description of the redshift evolution of the GCG energy density through the model parameter $\mathcal{A}_{c}$.
\vspace{0.2cm}\\
\hspace*{0.5cm} By substituting the energy density expression from Eq.~(\ref{20}) into the modified Friedmann equation in Eq.~(\ref{13}), together with the adopted form of the gravitational function $f(T)$ given in Eq.~(\ref{15}), we obtain the Hubble parameter as a function of redshift in the following form:
\begin{equation}{\label{21}}
	H^{2}(z) = \frac{\rho_{0}}{3(1-\gamma)} \left[\mathcal{A}_{c}+ (1-\mathcal{A}_{c})(1+z)^{3(1+\eta)}\right]^{\frac{1}{(1+\eta)}}.
\end{equation}
To establish a direct connection with the observational analysis, we define the present-day Hubble parameter $H_{0}$ through $H_{0}\equiv H(z=0)$. Evaluating Eq.~(\ref{21}) at the present epoch yields
\begin{equation}{\label{22}}
	H_{0} = \sqrt{\frac{\rho_{0}}{3(1-\gamma)} }.
\end{equation}
From this relation, the condition $\gamma<1$ follows from the requirement that $H_{0}^{2}$ remains positive. This constraint ensures a physically admissible background evolution within the present $f(T)$ framework. Using Eq.~(\ref{22}) to eliminate $\rho_{0}$, we can recast Eq.~(\ref{21}) as
\begin{equation}{\label{23}}
	H(z) = H_{0} \left[\mathcal{A}_{c}+ (1-\mathcal{A}_{c})(1+z)^{3(1+\eta)}\right]^{\frac{1}{2(1+\eta)}}.
\end{equation}
The Hubble function in Eq.~(\ref{23}) depends on the parameters $\mathcal{A}_{c}$, $\eta$ and $H_{0}$, which can be constrained by confronting the model predictions with cosmological observations.
\section{Observational constraints and statistical analysis}\label{sec:4}
\hspace*{0.5cm} In this section, we examine the compatibility of the proposed cosmological model with observational data by adopting a Bayesian parameter-estimation procedure. The model parameters $H_{0}$, $\mathcal{A}_{c}$ and $\eta$, which characterize the Hubble parameter in Eq.~(\ref{23}), are constrained using the cosmic chronometer (CC) measurements and the combined CC+Pantheon dataset. The parameter constraints are obtained through a Markov chain Monte Carlo (MCMC) analysis based on the $\chi^{2}$ statistic, with the sampling performed using the emcee Python package~\cite{foreman2013emcee}. The resulting constraints provide the preferred parameter values and their associated uncertainties, allowing us to assess the consistency of the proposed cosmological model with the considered observational datasets.
\subsection{The Cosmic chronometer dataset}\label{sec:4.1}
\begin{figure}[ht]
	\centering
	\includegraphics[width=11cm, height=6cm]{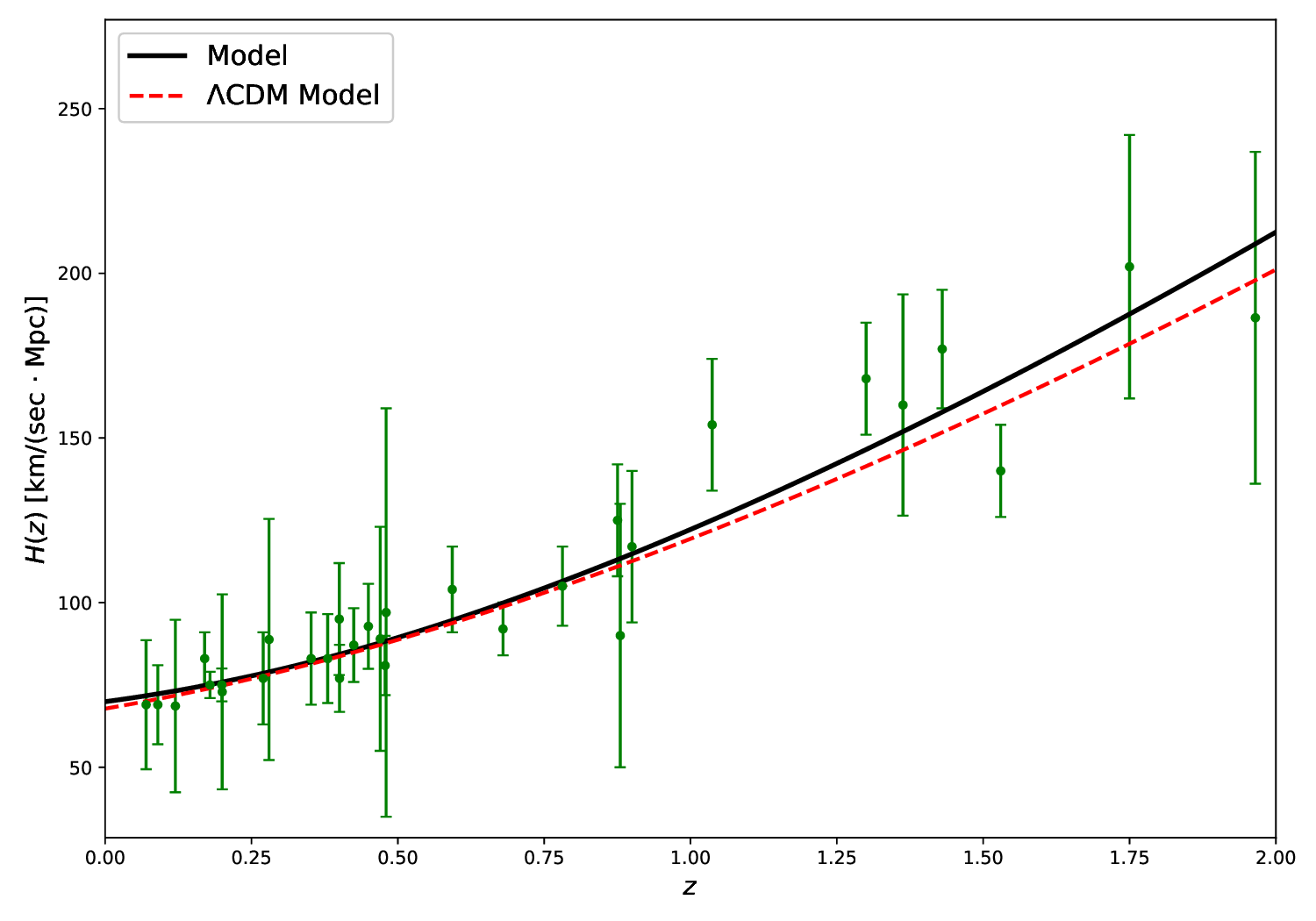}
	\caption{The best-fit evolution of the Hubble parameter $H(z)$ for the proposed model in comparison with the $\Lambda$CDM model.}
	\label{fig:1}
\end{figure}

\begin{figure}[htbp]
	\centering
	\includegraphics[width=10.0cm, height=10.0cm]{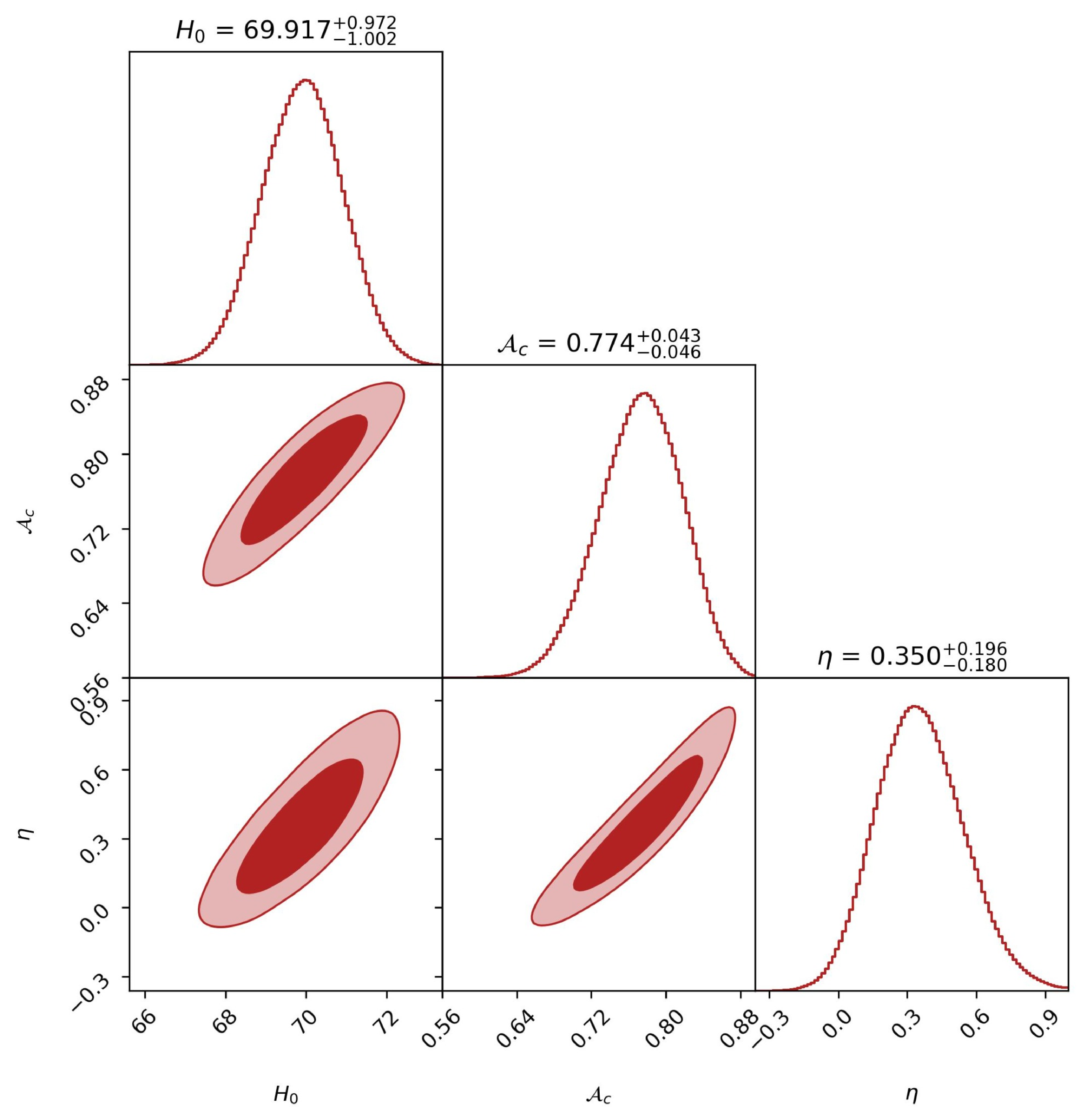}
	\caption{Marginalized $1D$ and $2D$ posterior contour map with median values of $H_{0}$, $\mathcal{A}_{c}$ and $\eta$ using the CC dataset.}
	\label{fig:2}
\end{figure}

\hspace*{0.5cm} To investigate the observational viability of the proposed cosmological scenario, we constrain its free parameters using observational measurements. We first consider a compilation of $31$ cosmic chronometer (CC) data points~\cite{simon2005constraints,sharov2018predictions}, obtained through the differential age (DA) method applied to passively evolving galaxies over the redshift range $0.07 \leq z \leq 1.965$~\cite{stern2010cosmic,moresco2015raising}. These measurements provide direct estimates of the Hubble expansion rate and therefore offer an independent means of constraining the background evolution. Following the method proposed by Jimenez and Loeb~\cite{jimenez2002constraining}, the Hubble parameter is related to the redshift and cosmic time according to $H(z)=-(1+z)^{-1}\frac{dz}{dt}$. The free parameters $H_{0}$, $\mathcal{A}_{c}$ and $\eta$ are constrained by minimizing the chi-squared $(\chi^{2})$ function (which corresponds to maximizing the associated likelihood function), given by~\cite{garg2025cosmological}:

\begin{equation}{\label{24}}
	\chi^{2}_{CC}(\theta)=\sum_{i=1}^{31} \frac{[H_{th}(\theta,z_{i})-H_{obs}(z_{i})]^{2} }{ \sigma^{2}_{H(z_{i})}},  
\end{equation} 
here, $H_{\mathrm{th}}$ denotes the theoretically predicted Hubble parameter, $H_{\mathrm{obs}}$ represents its observationally measured counterpart, and $\sigma_{H}$ corresponds to the associated observational uncertainty (standard deviation).
Figure~(\ref{fig:1}) displays the CC observational measurements along with their corresponding uncertainties and the best-fit curves of the Hubble parameter obtained from the analysis. Furthermore, Fig.~(\ref{fig:2}) presents the contour map corresponding to the $1\sigma$ and $2\sigma$ likelihood confidence levels for the median values of $H_{0}$, $\mathcal{A}_{c}$ and $\eta$ constrained from the CC dataset.
\subsection{The Pantheon dataset}\label{sec:4.2}
\begin{center}
	\begin{figure}
		\includegraphics[width=18.5cm, height=19cm]{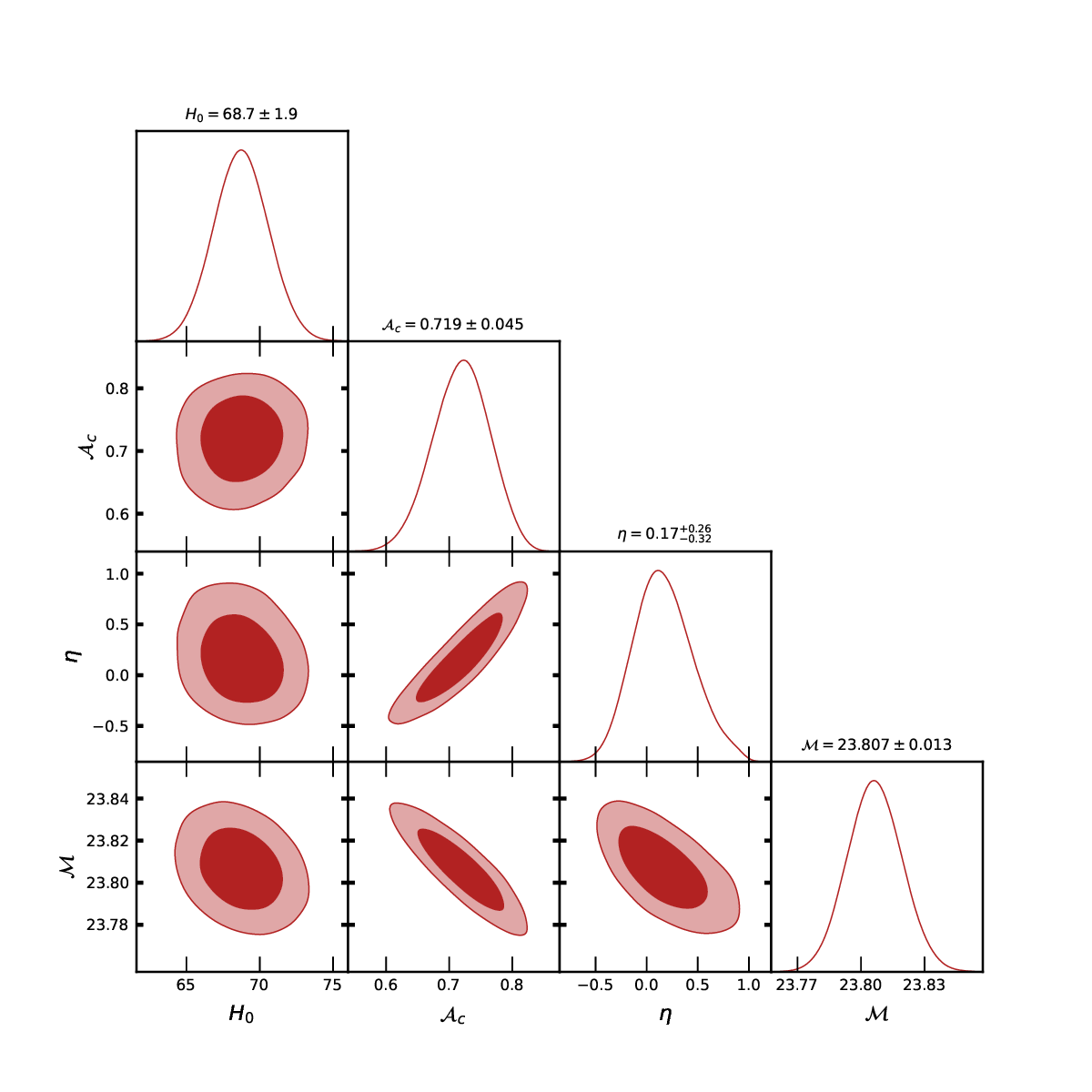}
		\caption{ Marginalized $1D$ and $2D$ posterior contour map with median values of $H_{0}$, $\mathcal{A}_{c}$ and $\eta$ using the Joint dataset.}
		\label{fig:3}
	\end{figure}
\end{center}

\hspace*{0.5cm} We further constrain the proposed model using the Pantheon compilation, which contains $1048$ Type Ia supernovae (SNe Ia) covering the redshift interval $0.01 < z < 2.26$~\cite{scolnic2018complete}. The compilation includes observations from several major surveys, namely the CfA1--CfA4 samples~\cite{riess1999bvri,hicken2009improved}, the Pan-STARRS1 Medium Deep Survey~\cite{scolnic2018complete}, the Sloan Digital Sky Survey (SDSS)~\cite{sako2018data}, the Supernova Legacy Survey (SNLS)~\cite{guy2010supernova}, and the Carnegie Supernova Project (CSP)~\cite{contreras2010carnegie}. For the Pantheon dataset, the theoretical apparent magnitude ($\mu_{\mathrm{th}}(z)$) used in the MCMC analysis can be written as
\begin{equation}{\label{25}}
\mu_{th}(z)=25+5log_{10}\left[\frac{d_{L}(z)}{Mpc}\right]+M,
\end{equation}
where $M$ denotes the absolute magnitude, while $d_{L}(z)$ represents the luminosity distance, which has dimensions of length and is given by~\cite{odintsov2018cosmological}
\begin{equation}{\label{26}}
	d_{L}(z)=c(1+z)\int_{0}^{z}\frac{dz'}{H(z')}.
\end{equation}
In this expression, $z$ denotes the redshift of the Type Ia supernovae (SNe Ia) measured in the cosmic microwave background (CMB) rest frame and $c$ is the speed of light. It is convenient to introduce the dimensionless, Hubble-free luminosity distance as $D_{L}(z) \equiv H_{0}d_{L}(z)/c$. Using this definition, Eq.~(\ref{25}) takes the following form:
\begin{equation}{\label{27}}
	\mu_{th}(z)=25+5log_{10}\left[D_{L}(z)\right]+5log_{10}\left[\frac{c/H_{0}}{Mpc}\right]+M. 
\end{equation}
The parameters $M$ and $H_{0}$ are not independently determined because of their well-established degeneracy in the $\Lambda$CDM framework~\cite{ellis2012relativistic,asvesta2022observational}. We therefore introduce the nuisance parameter $\mathcal{M}$ to absorb this degeneracy, which is defined as follows:

\begin{equation}{\label{28}}
	\mathcal{M}\equiv 25+5log_{10} \left[\frac{c/H_{0}}{Mpc}\right]+M=M+42.38-5log_{10}(h), 
\end{equation}
with $H_{0}=h \times 100$ $[\text{km}/(\text{sec}.\text{Mpc})]$, we conduct the MCMC analysis by simultaneously varying the relevant model parameters and employing the $\chi^{2}$ function associated with the Pantheon dataset, which is given by~\cite{asvesta2022observational,garg2025cosmological}:
\begin{equation}{\label{29}}
	\chi^{2}_{P}= \nabla \mu_{i}C^{-1}_{ij}\nabla \mu_{j}.
\end{equation}
Here $\nabla \mu_{i} = \mu_{obs}(z_{i}) - \mu_{th}(z_{i})$, with $C_{ij}^{-1}$ denoting the inverse covariance matrix and $\mu_{\mathrm{th}}$ specified by Eq.~(\ref{27}). Since the luminosity distance is determined by the expansion history encoded in the Hubble parameter, both observational datasets can be consistently incorporated into the parameter estimation. We use the emcee package~\cite{foreman2013emcee} to sample the parameter space within the adopted theoretical framework for the combined CC+Pantheon dataset. The corresponding joint statistic is obtained by adding the individual contributions, $\chi^{2}_{\mathrm{tot}}=\chi^{2}_{CC}+\chi^{2}_{P}$. Figure~(\ref{fig:3}) displays the resulting $1\sigma$ and $2\sigma$ confidence regions together with the marginalized $1D$ parameter distributions derived from the joint MCMC chains. The resulting median estimates of the model parameters from the MCMC analysis are reported in Table~(\ref{table:1}).
\begin{table}[htbp]
	\centering
	\renewcommand{\arraystretch}{2.5}  
	\fontsize{6pt}{9pt}\selectfont   
	\begin{tabular}{|c|c|c|c|c|c|c|c|c|c|c|c|}
			\hline
			Dataset & $H_{0}$[Km/(sec.Mpc)] & $\mathcal{A}_{c}$ & $\eta$ & $\mathcal{M}$ & $q_{0}$ & $z_{t}$ & $\omega_{0}$ & $j_{0}$ & $s_{0}$ & $t_{0}$(Gyr) \\
			\hline
			CC & $69.917^{+0.972}_{-1.002}$ & $0.774^{+0.043}_{-0.046}$ & $0.350^{+0.196}_{-0.180}$ & - & $-0.6610$ & $0.6125$ & $-0.7740$ & $1.2755$ & $-0.6331$ & $13.47^{+0.07}_{-0.06}$ \\
			\hline
			CC+Pantheon  & $68.7^{+1.9}_{-1.9}$  & $0.719^{+0.045}_{-0.045}$ &  $0.17^{+0.26}_{-0.32}$ &  $23.802^{+0.013}_{-0.013}$ & $-0.5785$ & $0.6114$ & $-0.7190$ & $1.1546$ & $-0.5430$ & $13.47^{+0.83}_{-0.45}$ \\
			\hline
	\end{tabular} 
\caption{Median values of the model parameters obtained from the CC and joint datasets, together with the present-day values of $q_{0}$, $\omega_{0}$, $j_{0}$, $s_{0}$ and $t_{0}$.}

	\label{table:1}
\end{table}
\vspace{0.2cm}\\
\hspace*{0.5cm} The MCMC analysis performed using the CC and joint (CC+Pantheon) observational datasets provides the constraints on the model parameters. For the CC dataset, the median values are $H_{0}=69.917^{+0.972}_{-1.002}$ $\mathrm{km s^{-1} Mpc^{-1}}$, $\mathcal{A}_{c}=0.774^{+0.043}_{-0.046}$ and $\eta=0.350^{+0.196}_{-0.180}$. For the joint dataset, the corresponding constraints are $H_{0}=68.7^{+1.9}_{-1.9}$ $\mathrm{km s^{-1} Mpc^{-1}}$, $\mathcal{A}_{c}=0.719^{+0.045}_{-0.045}$ and $\eta=0.17^{+0.26}_{-0.32}$. Furthermore, using the relation $H_{0}^{2}=\frac{\rho_{0}}{3(1-\gamma)}$ and adopting the normalization $\rho_{0}=1$, the observationally constrained values of $H_{0}$ yield $\gamma \approx 9.9993 \times 10^{-1}$ for both datasets.
\vspace{0.1cm}\\
\hspace*{0.5cm} The constrained values of the Hubble constant obtained from the CC and joint datasets are $H_{0}=69.917^{+0.972}_{-1.002}$ $\mathrm{km s^{-1} Mpc^{-1}}$ and $H_{0}=68.7^{+1.9}_{-1.9}$ $\mathrm{km s^{-1} Mpc^{-1}}$, respectively. These estimates lie between the local distance-ladder determination and the CMB-based inference within the standard $\Lambda$CDM framework, indicating that the present model yields an intermediate value of the Hubble expansion rate. These results provide an observationally motivated estimate of the present expansion rate and allow the model to be assessed in the context of the current Hubble-tension problem.
\vspace{0.1cm}\\
\hspace*{0.5cm} The parameter $\mathcal{A}_{c}$, which characterizes the relative contribution of the GCG component to the late-time cosmic evolution, is found to be approximately $0.77$ and $0.72$ for the CC and joint datasets, respectively. These values are close to the characteristic dark-energy contribution in the standard $\Lambda$CDM cosmology ($\Omega_{\Lambda,0}\sim0.7$), indicating that the GCG component can effectively account for the negative-pressure contribution associated with the late-time accelerated expansion within the present $f(T)$ framework. This result supports the viability of the GCG model in describing the observed accelerated expansion of the universe without introducing an explicit cosmological constant. 
\vspace{0.1cm}\\
\hspace*{0.5cm} The GCG parameter $\eta$ is constrained to be $0.35$ and $0.17$ for the CC and joint datasets, respectively. These constraints indicate that the GCG model allows a departure from the cosmological-constant limit while retaining its generalized fluid description. In particular, $\eta=0$ reduces the GCG equation of state to a constant negative pressure, corresponding to a cosmological-constant-like behavior, while $\eta=1$ recovers the standard Chaplygin gas. The constrained values of $\eta$ therefore indicate a moderate deviation from the cosmological-constant limit and support the generalized Chaplygin gas description of the late-time accelerated expansion.
\hspace*{0.5cm} Overall, the constrained values of $\mathcal{A}_{c}$ and $\eta$ demonstrate that the generalized Chaplygin gas within the $f(T)$ framework provides a unified description of the cosmic expansion history, encompassing matter-like behavior at early times and dark-energy-like behavior at late times. This unified description naturally accounts for the transition from decelerated to accelerated expansion without requiring an explicit cosmological constant. The consistency of the model parameters with the observational constraints supports the viability of the proposed framework as an alternative description of the late-time cosmic dynamics beyond the standard $\Lambda$CDM scenario.
\section{Evolution of cosmological parameters in $f(T)$ gravity}\label{sec:5}
\hspace*{0.5cm} In this section, we examine the evolution of the key cosmological quantities associated with the generalized Chaplygin gas scenario in the framework of $f(T)$ gravity. These quantities provide a quantitative characterization of the cosmic expansion history and offer useful diagnostics for gaining a better understanding of the dynamical behavior of the model across different stages of the universe's evolution. 
\subsection{Deceleration parameter}\label{sec:5.1}
\hspace*{0.5cm} The deceleration parameter $q(z)$ is an important quantity for describing the expansion dynamics of the universe. It is defined as $q(z)$=$-1-\frac{\dot{H}}{H^{2}}$, where $q>0$ represents a decelerating expansion, characteristic of the matter-dominated epoch, whereas $q<0$ signifies an accelerating expansion, as observed in the present epoch. Within the standard $\Lambda$CDM cosmology, the transition from decelerated to accelerated expansion is generally expected to take place around $z_{t}\sim0.6-0.8$, consistent with observational investigations of the late-time cosmic expansion. In the present study, we consider the Generalized Chaplygin Gas (GCG) model in the framework of $f(T)$ teleparallel gravity and obtain the corresponding expression for $q(z)$ from the Hubble parameter given in Eq.~(\ref{23}).
\begin{equation}{\label{30}}
	q(z) = -1+\frac{3}{2}\frac{ (1-\mathcal{A}_{c})(1+z)^{3(1+\eta)}}{\left[\mathcal{A}_{c}+ (1-\mathcal{A}_{c})(1+z)^{3(1+\eta)}\right]}.
\end{equation}
\begin{figure}[ht]
	\centering
	\includegraphics[width=10cm, height=6cm]{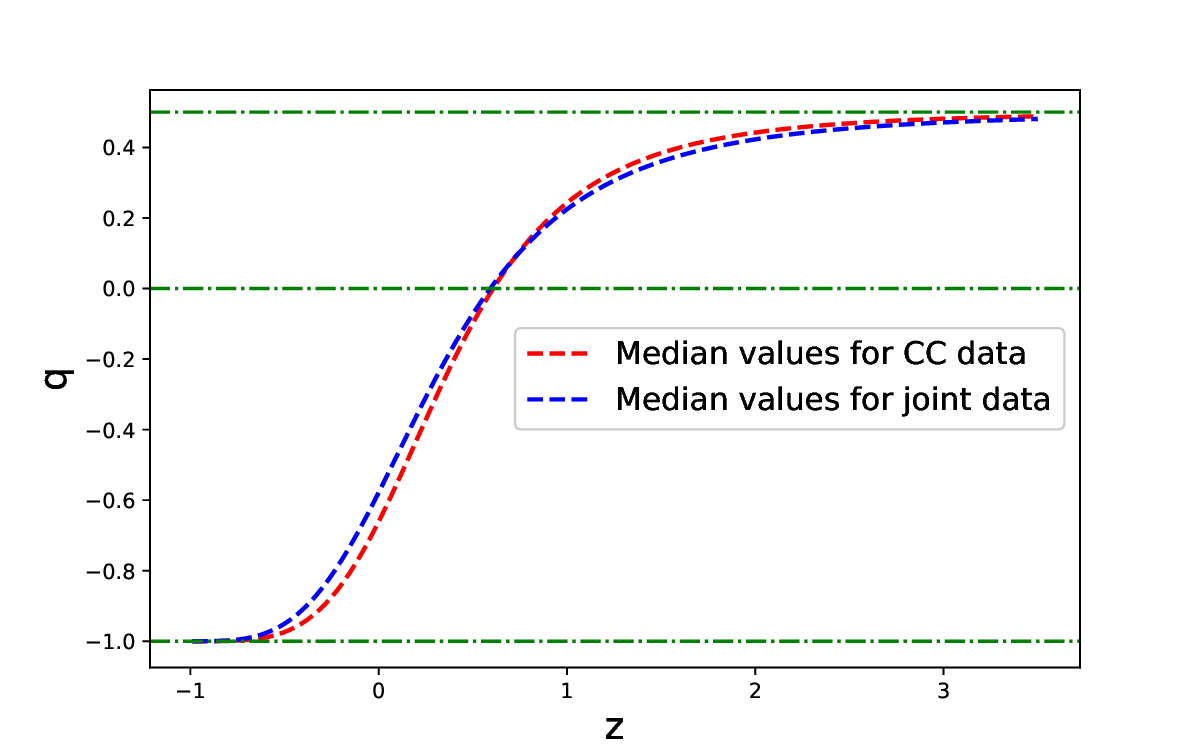}
	\caption{Profile of the $q(z)$ with redshift ($\mathit{z}$).}
	\label{fig:4}
\end{figure}
\hspace*{0.5cm} The redshift evolution of the deceleration parameter is illustrated in Fig.~(\ref{fig:4}). At large redshifts, $q(z)$ remains positive, showing that the universe was in a decelerating phase during its early evolution. This behavior is associated with the matter-dominated regime, which is crucial for the development of large-scale structures. As the redshift decreases, $q(z)$ gradually declines, reflecting the growing contribution of the GCG component to the cosmic dynamics. Consequently, the universe experiences a smooth transition from decelerated to accelerated expansion. From the observationally constrained model parameters, we obtain a transition redshift of $z_{t}=0.6125$ for the CC dataset, whereas the combined CC+Pantheon dataset yields $z_{t}=0.6114$ within the GCG+$f(T)$ framework. The close correspondence between these two transition redshifts indicates that the predicted cosmic transition remains consistent for both observational datasets. Furthermore, the resulting value $z_{t}\approx 0.61$ falls within the observationally expected range for the transition from deceleration to acceleration.
Beyond the transition epoch, $q(z)$ continues to decrease and tends towards $q(z)\rightarrow-1$ at late times, indicating a de Sitter-like accelerated expansion. At the present epoch ($z=0$), the model gives $q_{0}=-0.6610$ for the CC dataset and $q_{0}=-0.5785$ for the joint dataset. Since both values are negative, they confirm that the universe is presently undergoing accelerated expansion. Thus, the evolution of $q(z)$ exhibits a late-time behavior consistent with the accelerated expansion expected in the standard $\Lambda$CDM cosmology, while in the present model this acceleration arises naturally from the combined contribution of the GCG fluid and the modified teleparallel gravity framework.
\vspace{0.1cm}\\
\hspace*{0.5cm} The consistent behavior of the deceleration parameter $q(z)$ demonstrates that the proposed GCG+$f(T)$ framework is capable of describing the cosmic expansion history, beginning with the early decelerating phase and evolving towards the present-day accelerated phase. This evolution provides a natural explanation of the observed cosmic acceleration within the framework, without requiring a finely tuned cosmological constant.
\subsection{Evolution of energy density and pressure}\label{sec:5.2}
\hspace*{0.5cm} In this section, we examine the evolution of the energy density $\rho(z)$ and pressure $p(z)$ associated with the GCG within the framework of $f(T)$ teleparallel gravity. These quantities characterize the properties of the cosmic fluid and are essential for understanding the expansion dynamics of the universe. Using the expression for $\rho(z)$ obtained from the energy conservation equation and incorporating it into the modified field equations of $f(T)$ gravity, we derive the corresponding forms of $\rho(z)$ and $p(z)$. These expressions describe the evolution of the GCG fluid and its contribution to the cosmic expansion.
\begin{equation}{\label{31}}
	\rho(z) = 3H_{0}^{2} (1-\gamma) \left[\mathcal{A}_{c}+ (1-\mathcal{A}_{c})(1+z)^{3(1+\eta)}\right]^{\frac{1}{(1+\eta)}},
\end{equation}
\begin{equation}{\label{32}}
	p(z) = -3H_{0}^{2} (1-\gamma) \mathcal{A}_{c} \left[\mathcal{A}_{c}+ (1-\mathcal{A}_{c})(1+z)^{3(1+\eta)}\right]^{\frac{-\eta}{(1+\eta)}}.
\end{equation}
\begin{figure}[!htb]
	\captionsetup{skip=0.4\baselineskip,size=footnotesize}
	\begin{minipage}{0.50\textwidth}
		\centering
		\includegraphics[width=7cm,height=6cm]{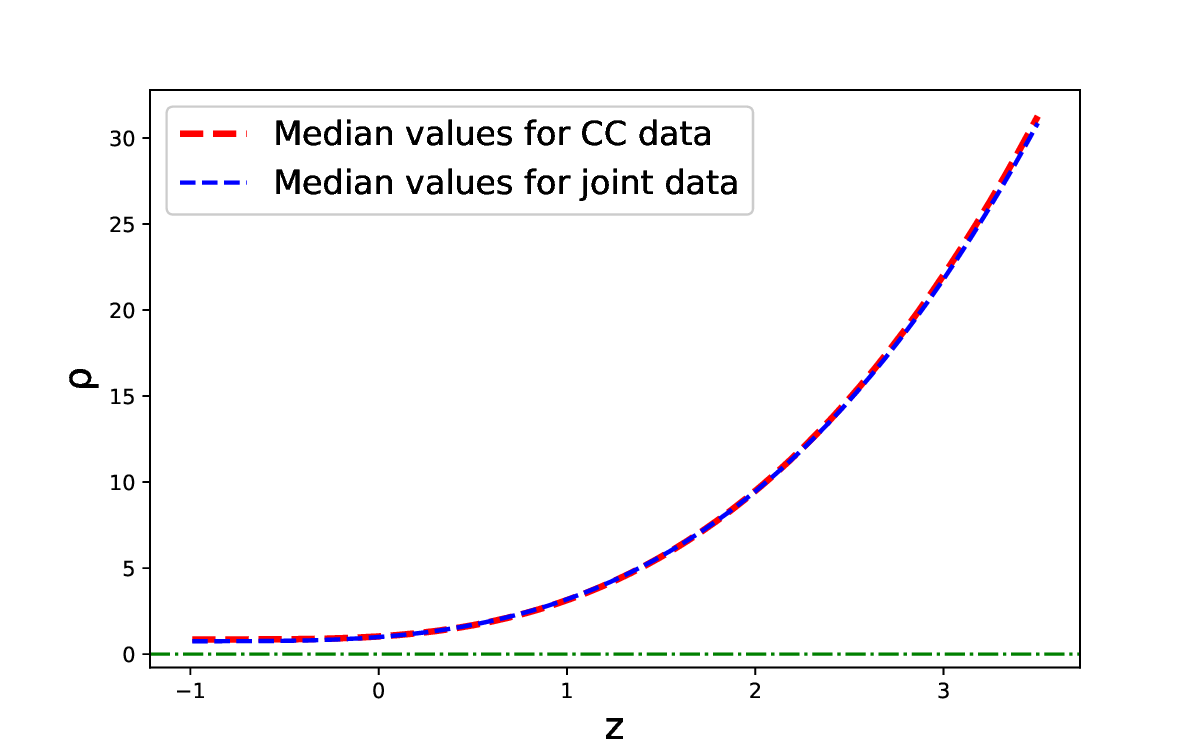}
		\caption{Profile of the energy density ($\rho$) with redshift ($\mathit{z}$).}
		\label{fig:5}
	\end{minipage}\hfill
	\begin{minipage}{0.50\textwidth}
		\centering
		\includegraphics[width=7cm,height=6cm]{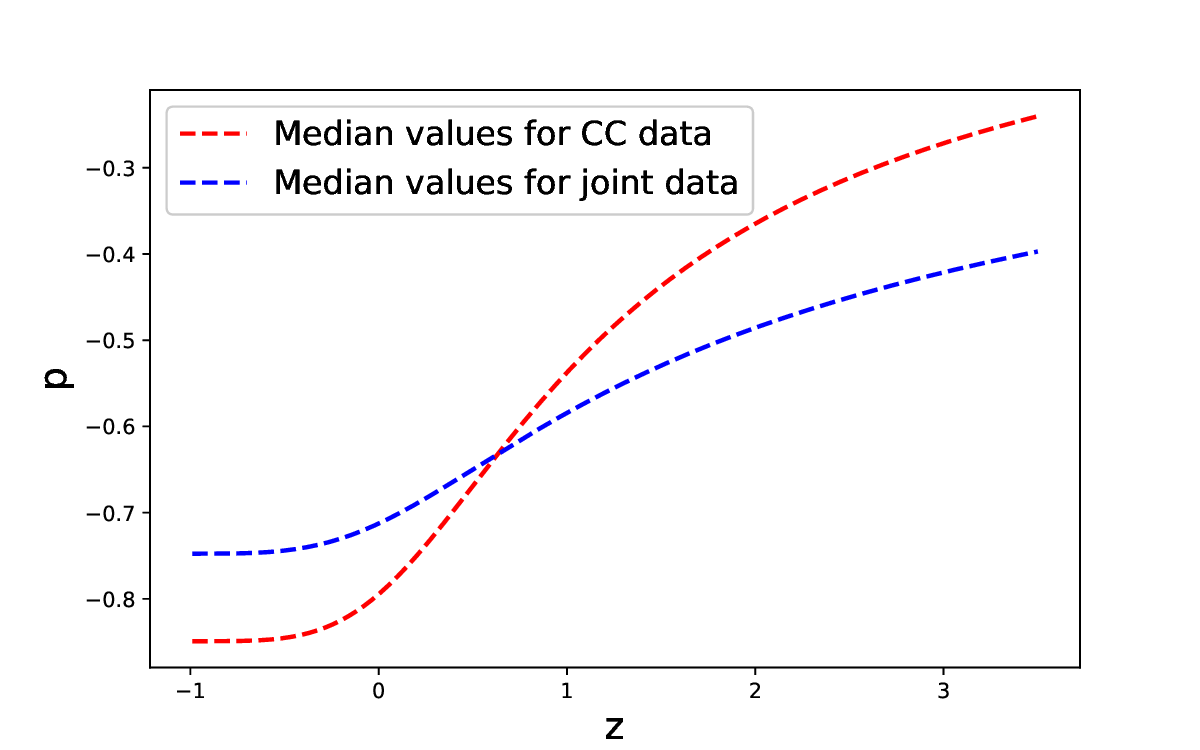}
		\caption{Profile of the pressure ($p$) with redshift ($\mathit{z}$).}
		\label{fig:6}
	\end{minipage}
\end{figure}
\hspace*{0.5cm} Figures~(\ref{fig:5}) and (\ref{fig:6}) present the redshift evolution of the energy density and pressure, respectively. The energy density remains positive over the considered redshift range, indicating the physical viability of the model and its consistency with the weak energy condition. On the other hand, the pressure becomes negative during the recent and late-time evolution of the universe, reflecting the dark-energy-like nature of the GCG component. This negative pressure contributes to the accelerated expansion of the universe and arises naturally from the dynamics of the GCG within the $f(T)$ teleparallel gravity framework.
\subsection{Evolution of the EoS parameter}\label{sec:5.3}
\hspace*{0.5cm} The Equation of State (EoS) parameter is an important quantity for describing the dynamical nature of the cosmic fluid and its role in the evolution of the universe. It is defined as the ratio of the pressure to the corresponding energy density $\left(\omega(z)=\frac{p(z)}{\rho(z)}\right)$ and provides a useful diagnostic for identifying the different regimes of cosmic evolution. In particular, $\omega=0$ corresponds to a pressureless matter-dominated universe, while $\omega=\frac{1}{3}$ represents the radiation-dominated epoch. On the other hand, $\omega=-1$ corresponds to a cosmological-constant or de Sitter-like state. The condition $\omega<-\frac{1}{3}$ is associated with accelerated cosmic expansion. Within this accelerating regime, the range $-1<\omega<-\frac{1}{3}$ corresponds to the quintessence-like behavior, whereas $\omega<-1$ represents the phantom regime, each leading to different possible dynamical behavior of the universe. In the present GCG+$f(T)$ framework, the redshift evolution of the EoS parameter $\omega(z)$ is obtained from the corresponding expressions of the energy density $\rho(z)$ and pressure $p(z)$ of our model. This evolution enables us to investigate the effective nature of the GCG component and to examine its contribution to the transition from the decelerated expansion at early times to the accelerated expansion at late times. The resulting behavior of $\omega(z)$ therefore provides an additional diagnostic for assessing the consistency of the proposed GCG+$f(T)$ model with the observed expansion history of the universe.
\begin{equation} {\label{33}}
\omega(z)= -1+ \frac{(1-\mathcal{A}_{c})(1+z)^{3(1+\eta)}}{\mathcal{A}_{c}+ (1-\mathcal{A}_{c})(1+z)^{3(1+\eta)}}.
\end{equation} 
\begin{figure}[ht]
	\centering
	\includegraphics[width=10cm, height=6cm]{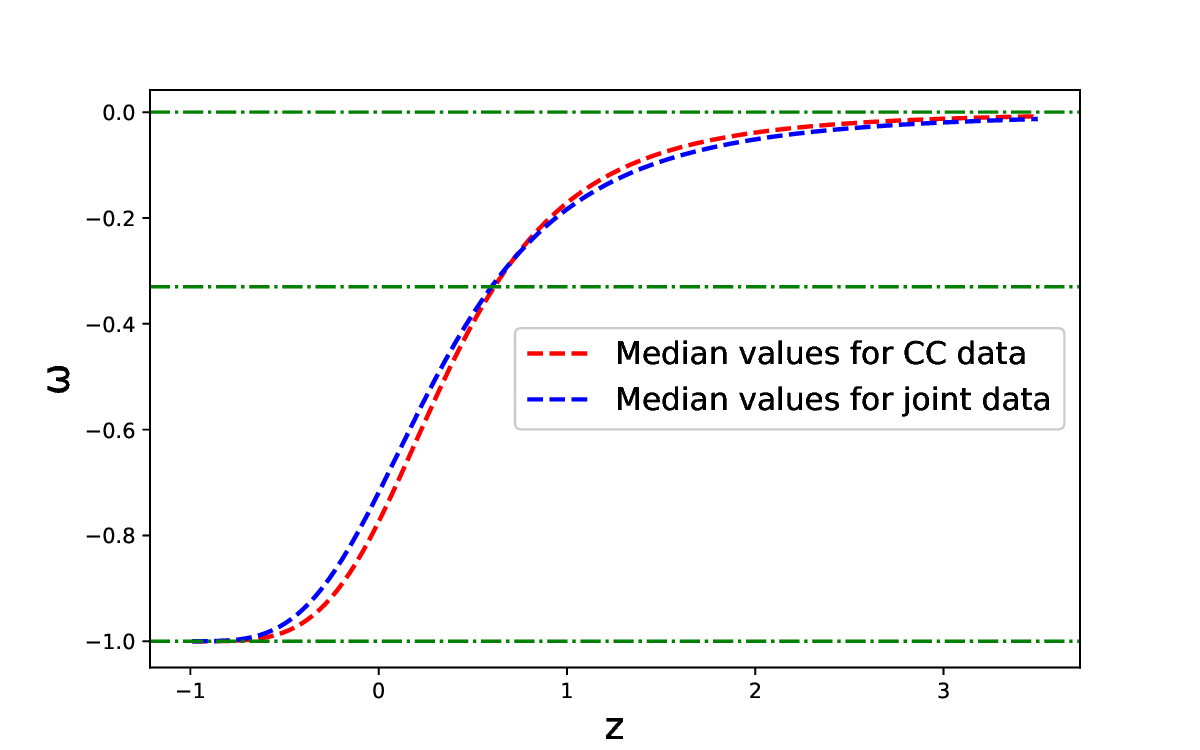}
	\caption{ Profile of the EoS parameter ($\omega$) with redshift ($\mathit{z}$).}
	\label{fig:7}
\end{figure}
\hspace*{0.5cm} The variation of $\omega(z)$ with redshift is displayed in Fig.~(\ref{fig:7}). The behavior of the EoS parameter provides a consistent picture of the cosmic evolution. At high redshifts $(z\gg1)$, the EoS parameter approaches a small positive value, $\omega(z\gg1)\approx0.0005$, indicating a nearly pressureless, matter-like behavior of the GCG component during the early universe. This behavior is compatible with the matter-dominated epoch and provides the appropriate conditions for the growth of cosmic structures. As the universe evolves towards lower redshifts, $\omega(z)$ gradually decreases and enters the negative regime, reflecting the increasing dark-energy-like contribution of the GCG component. This evolution accompanies the transition of the universe from decelerated to accelerated expansion.
\vspace{0.1cm}\\
\hspace*{0.5cm} In the low-redshift region $(z<1)$, the EoS parameter continues to decrease and reaches its present-day value at $z=0$. The model gives $\omega_{0}=-0.7740$ for the CC dataset and $\omega_{0}=-0.7190$ for the joint CC+Pantheon dataset. Both values lie in the quintessence regime ($-1<\omega_{0}<-\frac{1}{3}$), indicating a dark-energy-like behavior of the cosmic fluid at the present epoch. At late times $(z\rightarrow-1)$, the EoS parameter approaches $\omega(z)\rightarrow-1$, corresponding to a cosmological-constant-like behavior and a de Sitter-type accelerated phase. Hence, the evolution of $\omega(z)$ describes the major stages of cosmic evolution, beginning with an approximately matter-like early universe, followed by a smooth transition and ultimately approaching a dark-energy-dominated late-time state. This behavior demonstrates the capability of the GCG model in $f(T)$ teleparallel gravity to provide a unified description of cosmic evolution without invoking separate components.
\section{Energy conditions in $f(T)$ gravity with GCG}\label{sec:6}
\begin{figure}[ht]
	\centering
	\includegraphics[width=10cm, height=6cm]{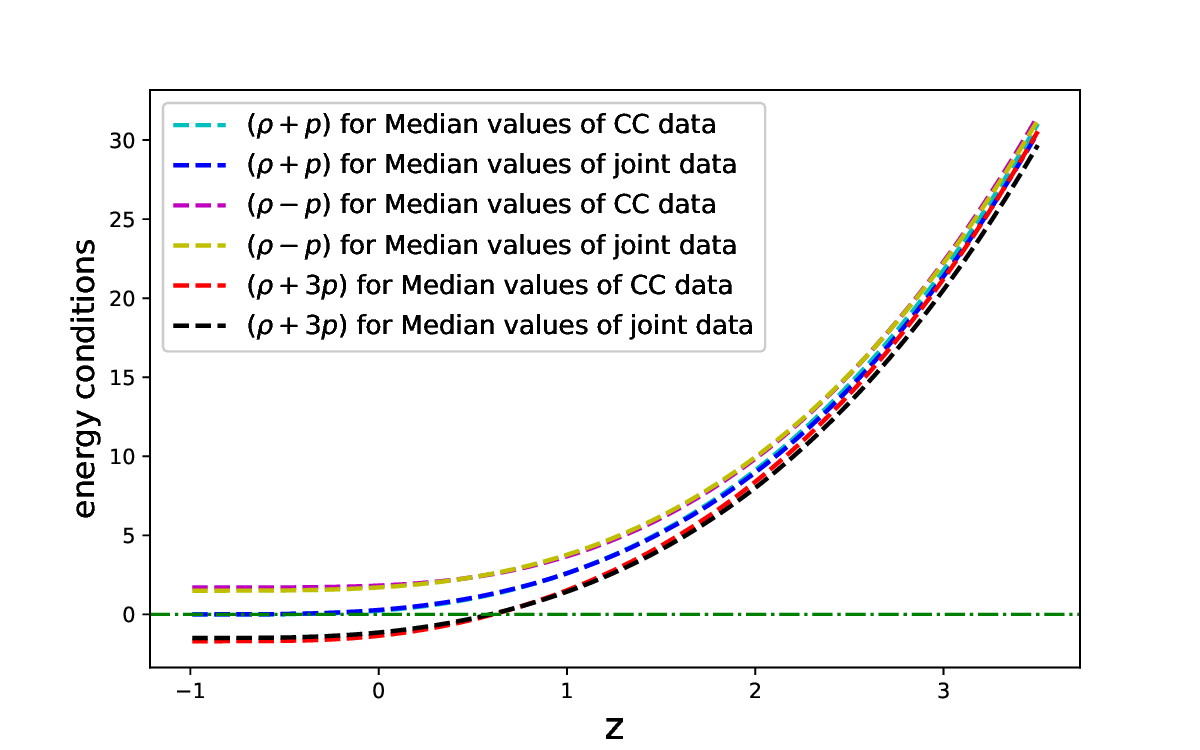}
	\caption{ Profile of the components of energy conditions with redshift ($\mathit{z}$).}
	\label{fig:8}
\end{figure}

\hspace*{0.5cm} Energy conditions provide a useful theoretical framework for examining the physical viability of cosmological models in general relativity and modified theories of gravity~\cite{visser1997energy,lalke2024cosmic,singh2022lagrangian}. They impose constraints on the energy-momentum content of the cosmic fluid and help determine whether the corresponding model satisfies fundamental physical requirements. In the present GCG model within $f(T)$ teleparallel gravity, we consider the standard Null Energy Condition (NEC), Dominant Energy Condition (DEC) and Strong Energy Condition (SEC), expressed as $\mathrm{NEC}:~\rho+p\geq0$, $\mathrm{DEC}:~\rho-p\geq0$ and $\mathrm{SEC}:~\rho+3p\geq0$, respectively. Using the derived expressions for the energy density $\rho(z)$ and pressure $p(z)$, we evaluate these energy conditions as functions of redshift. Their corresponding behaviors are displayed in Fig.~(\ref{fig:8}).
\begin{equation}{\label{34}}
	\rho +p = 3(1-\gamma) H_{0}^{2} (1-\mathcal{A}_{c})(1+z)^{3(1+\eta)} \left[\mathcal{A}_{c}+ (1-\mathcal{A}_{c})(1+z)^{3(1+\eta)}\right]^{\frac{-\eta}{(1+\eta)}},
\end{equation}
\begin{equation}{\label{35}}
	\rho -p = 3(1-\gamma) H_{0}^{2} \left[2\mathcal{A}_{c} + (1-\mathcal{A}_{c})(1+z)^{3(1+\eta)}\right] \left[\mathcal{A}_{c}+ (1-\mathcal{A}_{c})(1+z)^{3(1+\eta)}\right]^{\frac{-\eta}{(1+\eta)}},
\end{equation}
\begin{equation}{\label{36}}
	\rho +3p = 3(1-\gamma) H_{0}^{2} \left[(1-\mathcal{A}_{c})(1+z)^{3(1+\eta)}-2\mathcal{A}_{c}\right] \left[\mathcal{A}_{c}+ (1-\mathcal{A}_{c})(1+z)^{3(1+\eta)}\right]^{\frac{-\eta}{(1+\eta)}}.
\end{equation}
\hspace*{0.5cm} As shown in Fig.~(\ref{fig:8}), the NEC and DEC remain positive over the entire redshift range considered for both the CC and joint CC+Pantheon datasets. At high redshifts $(z\gg1)$, both conditions exhibit positive values, indicating that the cosmic fluid satisfies the corresponding energy requirements during the early stages of the universe. As the redshift decreases, the NEC and DEC gradually decline, reflecting the continuous evolution of the GCG component with cosmic expansion. Nevertheless, both quantities remain above their respective limiting values and do not show any violation throughout the considered evolution. Therefore, the persistence of the NEC and DEC indicates that the effective cosmic fluid remains physically well behaved within the proposed GCG+$f(T)$ framework.
The SEC displays a distinct behavior compared with the NEC and DEC. At high redshifts, the SEC remains positive, corresponding to the early decelerating phase of the universe. With decreasing redshift, the SEC gradually decreases and reaches zero at approximately $z\approx0.68$ for both the CC and joint datasets. Below this redshift, the SEC becomes negative and remains violated towards the present epoch and into the late-time regime. In particular, the violation persists at $z=0$, indicating that the present-day cosmic fluid does not satisfy the strong energy condition. This behavior is physically significant because the violation of $\rho+3p\geq0$ is associated with sufficiently negative pressure and permits accelerated expansion. Thus, the SEC evolution provides a clear indication of the change from the early decelerating regime, where the condition is satisfied, to the present and late-time accelerated regime, where the SEC is violated.
\vspace{0.1cm}\\
\hspace*{0.5cm} The simultaneous satisfaction of the NEC and DEC, and the transition of the SEC from a positive to a negative regime provide a consistent picture of the cosmic evolution in the GCG+$f(T)$ framework. The SEC violation beginning at approximately $z\approx0.68$ and continuing through the present epoch is consistent with the dark-energy-like behavior of the GCG component and the observed late-time accelerated expansion. The consistent behavior of the energy conditions for both datasets further supports the reliability of our results. Overall, the energy condition analysis supports the physical viability of the proposed GCG+$f(T)$ model and its ability to describe the evolution of the universe from the early decelerating phase to the present accelerated epoch.
\section{Dynamical analysis in the ($\omega-\omega'$) plane}\label{sec:7}
\hspace*{0.5cm} The $\omega-\omega'$ plane, introduced by Caldwell and Linder~\cite{caldwell2005limits}, is a useful diagnostic tool for investigating the dynamical behavior of dark-energy models. It provides a phase-space representation by plotting the equation of state parameter ($\omega$) against its derivative with respect to $\ln a$, defined as $\omega'=\frac{d\omega}{d\ln a}=-(1+z)\frac{d\omega}{dz}$. This representation allows different dynamical classes of dark-energy models, particularly the freezing ($\omega' < 0$ and $\omega < 0$) and thawing ($\omega' > 0$ and $\omega < 0$) behaviors, to be distinguished through their trajectories in the $\omega-\omega'$ plane. Using the expression obtained for $\omega(z)$ for the GCG model in the framework of $f(T)$ teleparallel gravity, we determine the corresponding trajectory in the $\omega-\omega'$ plane.
\begin{equation}{\label{37}}
	\omega' = -\frac{3(1+\eta) \mathcal{A}_{c} (1-\mathcal{A}_{c})(1+z)^{3(1+\eta)}}{\left[\mathcal{A}_{c}+ (1-\mathcal{A}_{c})(1+z)^{3(1+\eta)}\right]^{2}}.
\end{equation}
\begin{figure}[ht]
	\centering
	\includegraphics[width=10cm, height=5.5cm]{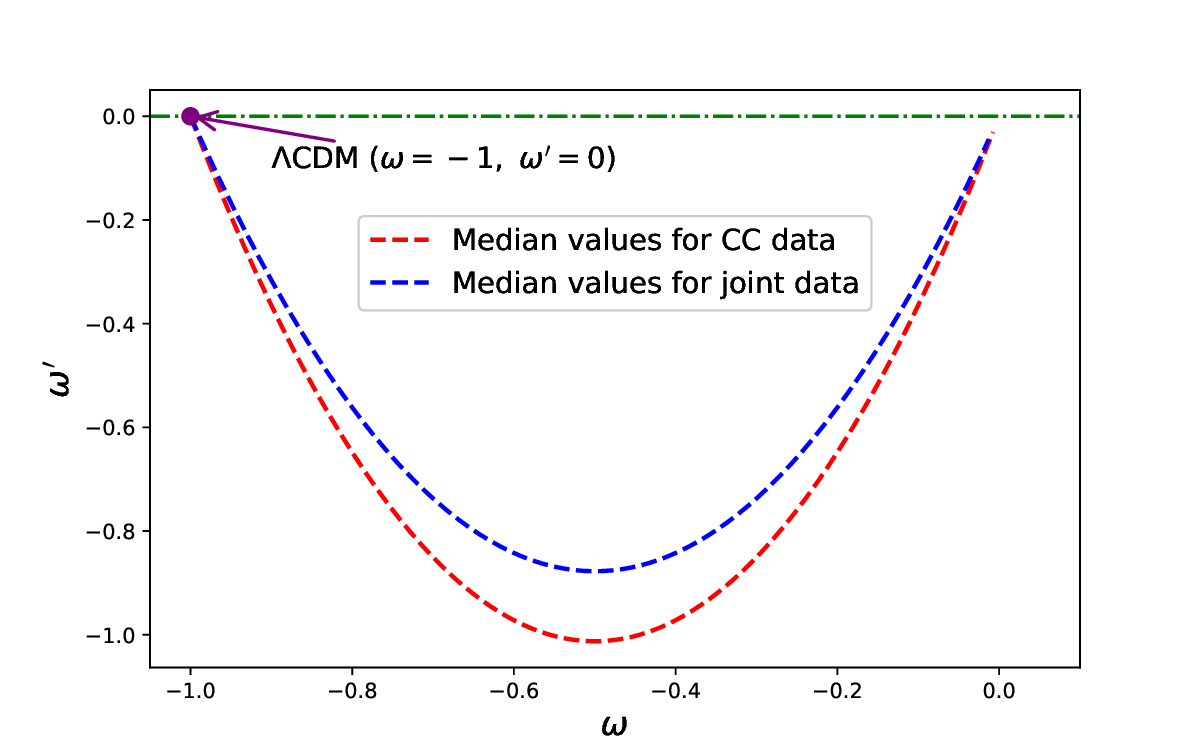}
	\caption{Profile of the ($\omega-\omega'$) plane.}
	\label{fig:9}
\end{figure}
\hspace*{0.5cm} Figure~(\ref{fig:9}) displays the corresponding evolutionary path. The trajectory initially lies close to the matter-like regime, with $(\omega,\omega')\approx(0.0012,-0.031)$, indicating that the GCG component behaves nearly like pressureless matter at early times with only a small rate of evolution. As the redshift decreases, the trajectory moves towards more negative values of $\omega$, while $\omega'$ undergoes a more pronounced variation. During this stage, the trajectory reaches a minimum around 
$(\omega,\omega')\approx(-0.5019,-1.013)$ for the CC dataset, whereas for the joint dataset, it reaches approximately 
$(\omega,\omega')\approx(-0.5019,-0.8793)$. 
These minimum points indicate a phase of enhanced dynamical evolution of the equation of state.
Following this stage, the trajectory enters the freezing region, where the variation of $\omega$ becomes progressively smaller and $\omega'$ moves towards zero. At late times, the trajectory approaches the fixed point $(\omega,\omega')=(-1,0)$, corresponding to the cosmological-constant or de Sitter limit. This asymptotic behavior indicates that the GCG component gradually approaches a cosmological-constant-like state in the far future, while also retaining dynamical properties during the recent past. The trajectory predominantly occupying the freezing region of the $\omega-\omega'$ plane further indicates a gradual reduction in the evolution of the EoS parameter towards the future. Thus, the $\omega-\omega'$ analysis provides an additional diagnostic of the dynamical behavior of the GCG+$f(T)$ framework and supports its ability to reproduce a cosmological-constant-like state at late times.
\section{Cosmographic analysis in $f(T)$ gravity with GCG: Jerk and Snap}\label{sec:8}
\begin{figure}[!htb]
	\captionsetup{skip=0.4\baselineskip,size=footnotesize}
	\begin{minipage}{0.50\textwidth}
		\centering
		\includegraphics[width=7cm,height=5.5cm]{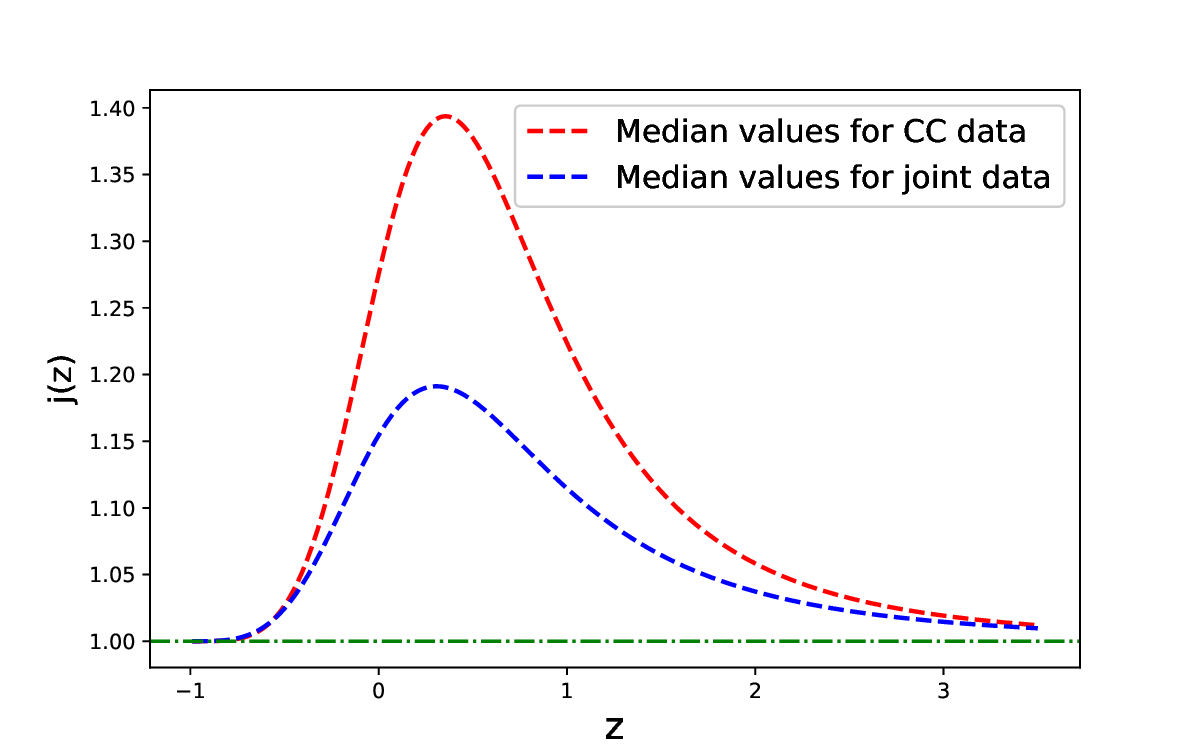}
		\caption{Profile of the jerk parameter ($\mathit{j}$) with redshift ($\mathit{z}$).}
		\label{fig:10}
	\end{minipage}\hfill
	\begin{minipage}{0.50\textwidth}
		\centering
		\includegraphics[width=7cm,height=5.5cm]{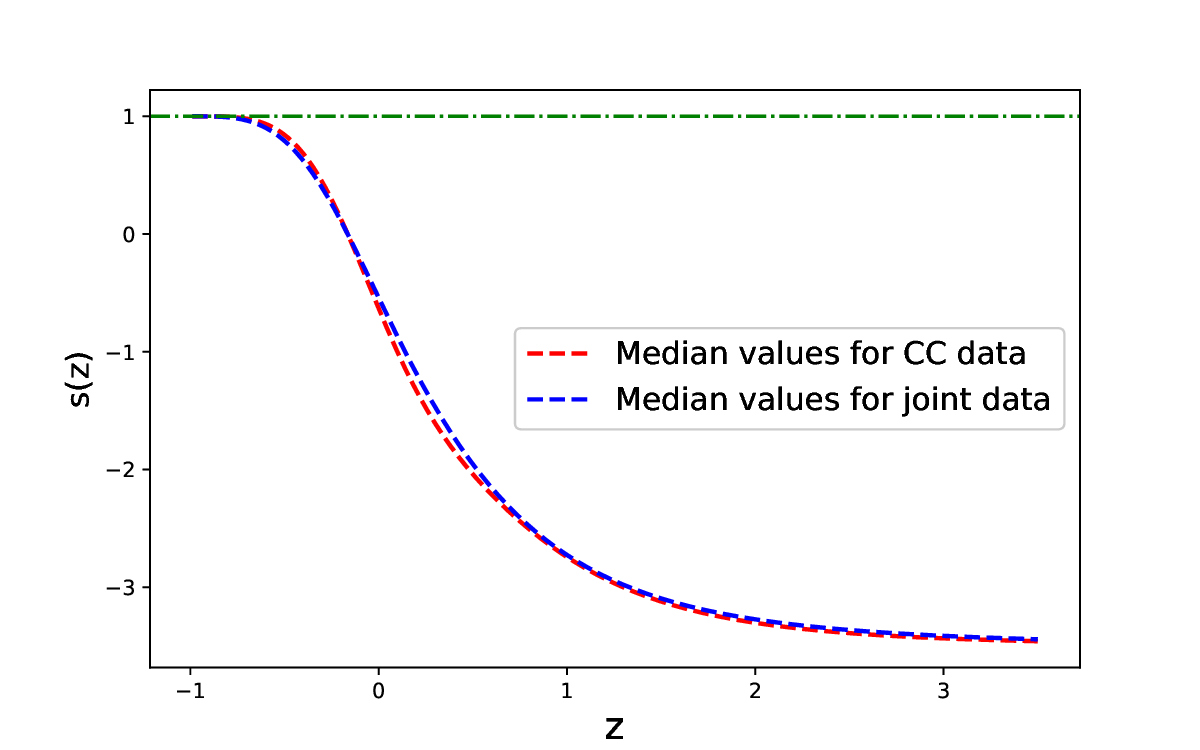}
		\caption{Profile of the snap parameter ($\mathit{s}$) with redshift ($\mathit{z}$).}
		\label{fig:11}
	\end{minipage}
\end{figure}

\hspace*{0.5cm} Cosmographic parameters provide a kinematic description of the expansion history of the universe through successive time derivatives of the scale factor. They are useful for examining the evolution of cosmic expansion without requiring a direct interpretation in terms of the underlying dynamical components. In particular, the jerk ($j$) and snap ($s$) parameters are associated with the third and fourth time derivatives of the scale factor, respectively, and characterize higher-order aspects of the cosmic expansion. These parameters therefore offer deeper insight into the kinematic evolution of the universe by probing higher-order variations in cosmic expansion and providing a refined perspective on its dynamical history. Following the cosmographic approach~\cite{visser2004jerk,wang2009probing}, the jerk and snap parameters are defined as
\begin{equation}{\label{38}}
	j=\frac{1}{aH^{3}}\frac{d^{3}a}{dt^{3}},
	\qquad
	s=\frac{1}{aH^{4}}\frac{d^{4}a}{dt^{4}}.
\end{equation}
\hspace*{0.5cm} To investigate the higher-order kinematic behavior of the GCG+$f(T)$ model, the jerk and snap parameters are expressed in terms of the redshift-dependent deceleration parameter as
\begin{equation}{\label{39}}
	j(z)=q(z)\left[1+2q(z)\right]+(1+z)\frac{dq}{dz},
\end{equation}
\begin{equation}{\label{40}}
	s(z)=-j(z)\left[2+3q(z)\right]-(1+z)\frac{dj}{dz}.
\end{equation}
\hspace*{0.5cm} Using the expression of $q(z)$ obtained for the GCG+$f(T)$ framework into these relations, we derive the corresponding analytical expressions for $j(z)$ and $s(z)$. Their respective redshift evolutions are shown in Figs.~(\ref{fig:10}) and~(\ref{fig:11}).
\begin{equation}{\label{41}}
	j(z) = 1+\frac{9}{2}\eta \frac{\mathcal{A}_{c} (1-\mathcal{A}_{c})(1+z)^{3(1+\eta)}}{\left[\mathcal{A}_{c}+ (1-\mathcal{A}_{c})(1+z)^{3(1+\eta)}\right]^{2}},
\end{equation}
\begin{equation} {\label{42}}
	\resizebox{1.0\textwidth}{!}{$	
		s(z) = 1-\frac{9}{2} \frac{ (1-\mathcal{A}_{c})(1+z)^{3(1+\eta)}}{\left[\mathcal{A}_{c}+ (1-\mathcal{A}_{c})(1+z)^{3(1+\eta)}\right]} + \frac{9}{2} \eta \frac{ \mathcal{A}_{c} (1-\mathcal{A}_{c})(1+z)^{3(1+\eta)}}{\left[\mathcal{A}_{c}+ (1-\mathcal{A}_{c})(1+z)^{3(1+\eta)}\right]^{2}} \left\{-2-3\eta+\left(\frac{3}{2}+6\eta\right)\frac{ (1-\mathcal{A}_{c})(1+z)^{3(1+\eta)}}{\left[\mathcal{A}_{c}+ (1-\mathcal{A}_{c})(1+z)^{3(1+\eta)}\right]}\right\}
		$}.
\end{equation} 
\hspace*{0.5cm} Figures~(\ref{fig:10}) and~(\ref{fig:11}) illustrate the redshift evolution of the jerk and snap parameters, respectively, obtained using the median values of the model parameters for the CC and joint CC+Pantheon datasets. For the CC dataset, the present-day values of the jerk and snap parameters are found to be $\mathit{j_{0}}=1.2755$ and $\mathit{s_{0}}=-0.6331$, respectively. The corresponding values for the joint dataset are $\mathit{j_{0}}=1.1546$ and $\mathit{s_{0}}=-0.5430$. In the standard $\Lambda$CDM framework, the present-day jerk parameter is expected to take the value $j_{0}=1$, whereas the evolution of the snap parameter depends on the underlying cosmological parameters. The obtained values of $j_{0}$ indicate a moderate departure from the $\Lambda$CDM prediction, with the joint result showing a closer correspondence to the standard scenario than the CC result.
\vspace{0.1cm}\\
\hspace*{0.5cm} The redshift evolution of the jerk and snap parameters provides further insight into the higher-order kinematic behavior of the GCG+$f(T)$ framework. At higher redshifts, the evolution of these parameters exhibits a noticeable departure from the corresponding $\Lambda$CDM behavior, which can be associated with the dynamical nature of the GCG component during the early stages of cosmic evolution. As the universe evolves towards the present epoch and subsequently into the late-time regime, both cosmographic parameters gradually approach the behavior expected for a $\Lambda$CDM-like expansion. This indicates that the proposed model can reproduce a $\Lambda$CDM-like behavior at late times while allowing for modified kinematic evolution at earlier epochs. Overall, the evolution of the jerk and snap parameters highlights the kinematic consistency of the GCG+$f(T)$ framework and further supports its capability to describe the late-time cosmic expansion. 
\section{Age of the Universe}\label{sec:9}
\hspace*{0.5cm} The cosmic age $t(z)$ provides a useful measure of the temporal evolution of the universe and can be expressed as a function of redshift $z$~\cite{tong2009cosmic}:
\begin{equation}{\label{43}}
	t(z)=\int_{z}^{\infty}\frac{dz}{(1+z)H(z)}.
\end{equation}
The present age of the universe ($t_{0}$), is obtained by evaluating the above integral at the present epoch ($z=0$), using the Hubble parameter (from Eq.~(\ref{23})) corresponding to the GCG+$f(T)$ framework. For the CC dataset, we obtain $t_{0}=13.47^{+0.07}_{-0.06}$ Gyr, whereas the joint CC+Pantheon dataset gives $t_{0}=13.47^{+0.83}_{-0.45}$ Gyr. Both estimates are reasonably close to the standard $\Lambda$CDM estimate of approximately $13.8$ Gyr~\cite{2020A&A...641A...6P}, indicating that the proposed framework yields a cosmic age compatible with the observationally inferred age of the universe. The close agreement between the results obtained from the two observational datasets further supports the viability of the GCG+$f(T)$ framework in describing the late-time evolution of the universe.
\section{Conclusions}\label{sec:10}
In this work, we have investigated the cosmological dynamics of the universe within the framework of $f(T)$ teleparallel gravity, where the gravitational Lagrangian is taken as $f(T)=\beta(-T)^{1/2}+\gamma(-T)$. To describe the cosmic fluid, we have employed the Generalized Chaplygin Gas (GCG) equation of state $p=-\frac{\mathcal{A}}{\rho^{\eta}}$, which provides a unified description of the cosmic evolution from a matter-like phase to a dark-energy-like phase. Within this framework, we derived an explicit form of the Hubble parameter $H(z)$ as a function of redshift. The reconstructed $H(z)$ profiles presented in Fig.~(\ref{fig:1}) show good agreement with the cosmic chronometer observations, supporting the consistency of the model with the observed expansion history.
\begin{itemize}
\item Using an MCMC-based statistical analysis, we constrained the model parameters using the CC and joint (CC+Pantheon) observational datasets. For the CC dataset, the constrained values are $H_{0}=69.917^{+0.972}_{-1.002}$ km s$^{-1}$ Mpc$^{-1}$, $\mathcal{A}_{c}=0.774^{+0.043}_{-0.046}$ and $\eta=0.350^{+0.196}_{-0.180}$, whereas the joint dataset gives $H_{0}=68.7^{+1.9}_{-1.9}$ km s$^{-1}$ Mpc$^{-1}$, $\mathcal{A}_{c}=0.719^{+0.045}_{-0.045}$ and $\eta=0.17^{+0.26}_{-0.32}$. These results provide observational constraints on the expansion history and demonstrate the compatibility of the proposed framework with the considered cosmological datasets.
\item We then studied the behavior of several cosmological parameters. The deceleration parameter $q(z)$ (in Fig.~(\ref{fig:4})) shows a transition from deceleration to acceleration at $z_{t}=0.6125$ for the CC dataset and $z_{t}=0.6114$ for the joint dataset, with present-day values $q_{0}=-0.6610$ and $q_{0}=-0.5785$, respectively. These results confirm the accelerated expansion of the universe at the present epoch and indicate a consistent transition redshift for both observational datasets.
\item The evolution of the energy density, pressure and EoS parameter further supports the cosmological behavior of the model. The energy density remains positive over the considered redshift range, whereas the pressure becomes negative at late times. The EoS parameter (in Fig.~(\ref{fig:7})) begins near zero at early times
(indicating matter-like behavior), decreases towards negative values during the subsequent evolution and reaches $\omega_{0}=-0.7740$ for the CC dataset and $\omega_{0}=-0.7190$ for the joint dataset at the present epoch. In the asymptotic future, it approaches $\omega=-1$, corresponding to a cosmological-constant-like state.
\item The model was also tested through the energy conditions (in Fig.~(\ref{fig:8})). We found that the NEC and DEC remain satisfied throughout the cosmic evolution, whereas the SEC becomes violated at approximately $z\approx0.68$. This late-time SEC violation is associated with the accelerated expansion of the universe and is consistent with the negative pressure behavior of the GCG component.
\item We further explored the dynamical behavior in the $\omega-\omega'$ plane (in Fig.~(\ref{fig:9})). The trajectory starts close to the matter-like region, evolves through an intermediate dynamical phase and finally enters the freezing region, approaching the fixed point $(\omega,\omega')=(-1,0)$ at late times. This behavior indicates an asymptotic evolution towards a cosmological-constant-like state. In addition, the present-day jerk and snap parameters are found to be $j_{0}=1.2755$, $s_{0}=-0.6331$ for the CC dataset and $j_{0}=1.1546$, $s_{0}=-0.5430$ for the joint dataset, providing further information about the higher-order expansion dynamics (in Figs.~(\ref{fig:10}) to (\ref{fig:11})).
\item Finally, we calculated the age of the universe using the Hubble parameter obtained from the proposed GCG+$f(T)$ framework. The present age is found to be $t_{0}=13.47^{+0.07}_{-0.06}$ Gyr for the CC dataset and $t_{0}=13.47^{+0.83}_{-0.45}$ Gyr for the joint CC+Pantheon dataset. These values are reasonably close to the standard $\Lambda$CDM estimate, indicating that the proposed model provides a consistent estimate of the age of the universe.
\end{itemize}
In summary, the GCG model embedded in $f(T)$ teleparallel gravity provides a unified description of cosmic evolution, connecting the early matter-like decelerating phase with the present dark-energy-like accelerated epoch. The observational constraints, evolution of the cosmological parameters, satisfaction of the NEC and DEC, late-time SEC violation, freezing behavior in the $\omega-\omega'$ plane, cosmographic diagnostics, and estimated cosmic age collectively support the viability of the proposed framework. The model therefore provides an alternative description of the late-time accelerated expansion of the universe without introducing an explicit cosmological constant, while retaining a close correspondence with the standard $\Lambda$CDM behavior at late times, thereby offering a viable framework for understanding cosmic acceleration within the context of modified gravity.
\section*{\textbf{Acknowledgements}}
GPS is thankful to the Inter-University Centre for Astronomy and Astrophysics (IUCAA), Pune, India for support under Visiting Associateship program.

\end{document}